\documentclass{article}
\usepackage{authblk}
\usepackage{arxiv}
\usepackage[utf8]{inputenc} 
\usepackage[T1]{fontenc}    
\usepackage{hyperref}       
\usepackage{url}            
\usepackage{booktabs}       
\usepackage{amsfonts}       
\usepackage{nicefrac}       
\usepackage{microtype}      
\usepackage{graphicx}
\usepackage{caption}
\usepackage{subfigure}

\title{Substituting Gadolinium in Brain MRI Using DeepContrast}
\author[*1]{Haoran Sun}
\author[*1]{Xueqing Liu}
\author[1]{Xinyang Feng}
\author[2]{Chen Liu}
\author[3]{Nanyan Zhu} 
\author[1]{Sabrina J. Gjerswold-Selleck}
\author[8,10]{\\Hongjian Wei}
\author[9,10]{Pavan S. Upadhyayula}
\author[9,10]{Angeliki Mela}
\author[8,10]{Chengchia Wu}
\author[9,10]{Peter D. Canoll}
\author[1]{\\Andrew F. Laine}
\author[4,6,7]{Scott A. Small}
\author[4,5]{Jia Guo}
\affil[1]{Department of Biomedical Engineering,$^2$Department of Electrical Engineering, $^3$Department of Biological Science,}
\affil[4]{Departments of Psychiatry, $^5$Zuckerman Institute, $^6$Departments of Neurology,$^7$Departments of Radiology,}
\affil[8]{Department of Radiation Oncology, and $^9$Department of Pathology and Cell Biology, Columbia University, New York, NY.
}

\affil[10]{Columbia University Irving Medical Center, New York, NY.}

\date{} 

\begin{document}
\footnote{* These authors contributed equally to this work.}

\footnote{Correspondence: Jia Guo  (jg3400@columbia.edu)}

\footnote{Accepted for publication at ISBI 2020. 
\copyright 2020 IEEE
}

\maketitle

\begin{abstract}
Cerebral blood volume (CBV) is a hemodynamic correlate of oxygen metabolism and reflects brain activity and function. High-resolution CBV maps can be generated using the steady-state gadolinium-enhanced MRI technique. Such technique requires an intravenous injection of exogenous gadolinium based contrast agent (GBCA) and recent studies suggest that the GBCA can accumulate in the brain after frequent use. We hypothesize that endogenous sources of contrast might exist within the most conventional and commonly acquired structural MRI, potentially obviating the need for exogenous contrast. Here, we test this hypothesis by developing and optimizing a deep learning algorithm, which we call DeepContrast, in mice. We find that DeepContrast performs equally well as exogenous GBCA in mapping CBV of the normal brain tissue and enhancing glioblastoma.  Together, these studies validate our hypothesis that a deep learning approach can potentially replace the need for GBCAs in brain MRI. 
\end{abstract}


\section{Introduction}
Information extracted from MRI can be dramatically enhanced with the use of the exogenous contrast agent, gadolinium (Gd). Gd-enhanced MRI is routinely used to better visualize nearly all neurological disease - including strokes, tumors, infections, and neuroinflammation. Moreover, since steady-state Gd-enhanced MRI technique can generate high resolution maps of the cerebral blood volume (CBV) and cerebral blood flow (CBF), both tightly coupled to brain metabolism, Gd-enhanced MRI can be used as an fMRI tool \cite{RN10}. More recently, because it generates CBV maps that are both quantitative and have high spatial resolution, this CBV-fMRI approach has been used to detect the earliest stages of Alzheimer's disease \cite{RN12} and schizophrenia \cite{RN11}, and to map the effects of normal aging \cite{RN16}.

Despite its significant advantages, reports of gadolinium retention in the brain and body after previous exposure to gadolinium based contrast agents (GBCAs) raise serious safety concerns in the clinical community \cite{RN5}. Given the potential safety risks and the FDA warning \cite{RN40}, there is an urgent need to develop alternative imaging techniques that reduce the dose of Gd or eliminate it entirely to prevent Gd retention.

Recently, deep learning has shown a great potential to reduce Gd exposure in brain MRI \cite{Gong}. This study shows that Gd dose can be reduced 10-fold while preserving contrast information and avoiding significant image quality degradation for visibility of pathology and delineation of brain lesions \cite{Gong}. A recent study further demonstrates the possibility of predicting virtual Gd contrast in brain MRI from noncontrast multiparametric MRI scans \cite{Can}. The question remains whether it is feasible to estimate Gd contrast in brain MRI directly from a single noncontrast scan such as the most readily available structural MRI to ensure patient safety and imaging efficiency. We hypothesize that deep learning can produce Gd contrast in brain MRI directly from single noncontrast structural MRI. In the current study, we test this idea in mice as a proof of concept study before translating the same approach to humans. We utilize the residual attention U-Net architecture in our deep learning model to estimate Gd contrast from noncontrast T2-weighted (T2W) MRI for CBV mapping. Our method is evaluated for both wild-type (WT) mice and mice with Glioblastoma (GBM) at 9.4T.

\section{MATERIAL AND METHODS}
\label{sec:headings}
Experiments were performed following the National Institutes of Health (NIH) guidelines and were approved by the Institutional Animal Care and Use Committee (IACUC).

\subsection{Animal Subject}
Mice used in our study were divided into two groups: WT mice and mice with GBM. The WT group contained 49 healthy adult C576J/BL male mice scanned at 12-14 months old. The GBM group contained 10 adult C576J/BL male mice that were injected with PDGFB (+/+) PTEN (-/-) p53 (-/-) glioblastoma (GBM) cells \cite{Lei}. 50,000 cells in 1 $\mu L$ were stereotactically injected into the brain. MRI scans of GBM mice were obtained 10 days after injection.

\begin{figure}
  \centering
  \includegraphics[width=0.8\textwidth]{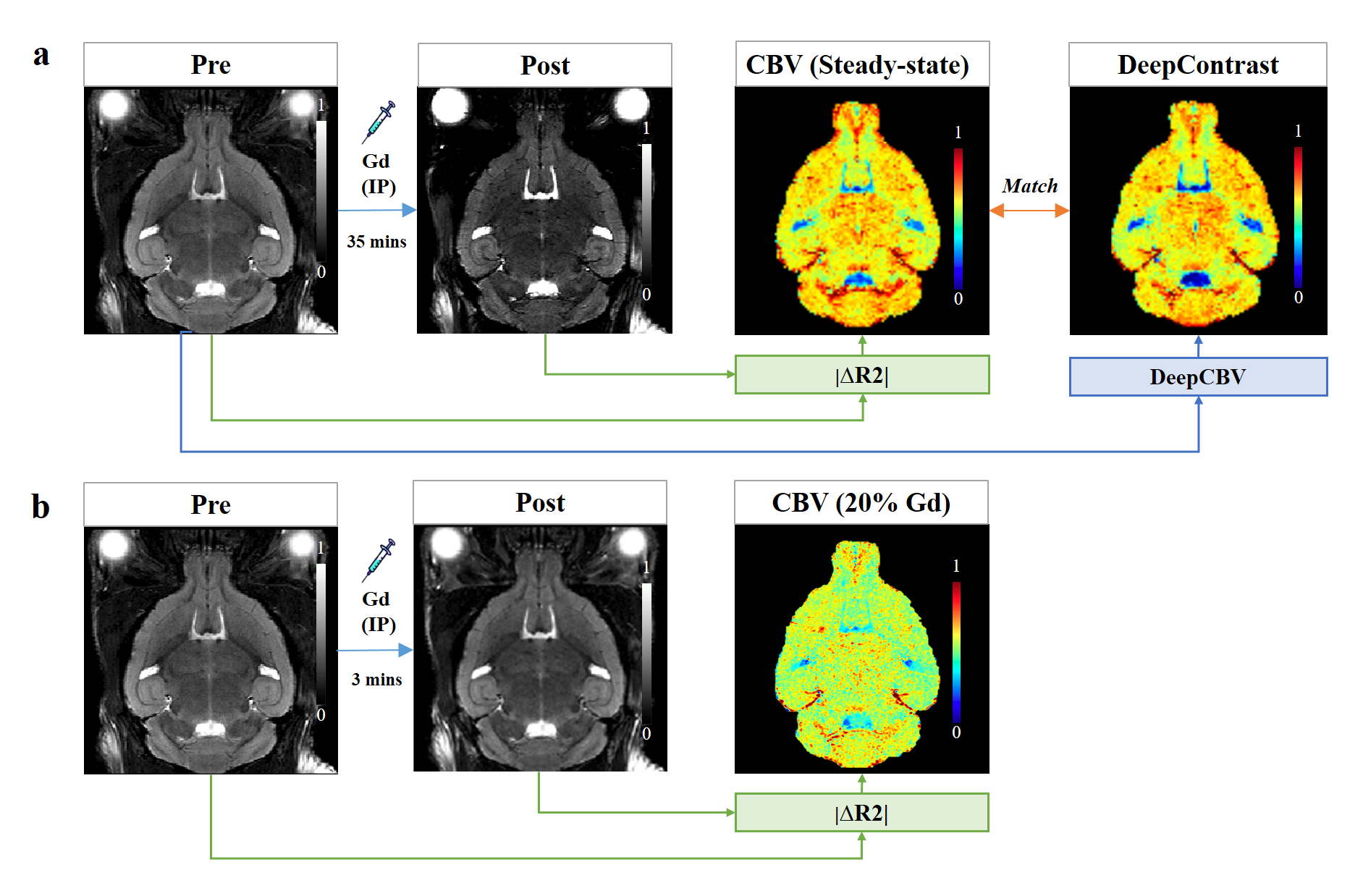} 
  \caption{The T2 weighted MRI acquisition pipeline and the generation of $\Delta$R2 map. a) The steady-state CBV map was derived by the T2W MRI Pre scans before the IP injection and the Post scans 35 mins after the IP injection. The DeepContrast map match the CBV map. b) As a direct approach to reduce Gd dose, the low dose CBV map can be generated by the Pre scans and the Post scans 3 mins after the IP injection.}
  \label{Figure:Figure1}
\end{figure}

\subsection{MRI Acquisition and Preprocessing}
For each mouse, T2W MRI scans were acquired using the 2D T2-weighted Turbo Rapid Acquisition with Refocused Echoes (RARE) sequence at 9.4T (i.e., TR/TE = 3500/45, RARE factor = 8, 76 $\mu m$ in-plane resolution, 450 $\mu m$ slice thickness; Bruker Biospec 94/30 USR equipped with CryoProbe). 

As previously described, to derive the steady-state CBV maps, whole brain T2W MRI scans before (i.e., Pre) and 35 mins after (i.e., Post) IP injection of Gadodiamide at 10 $mmol/kg$ are acquired with identical scan parameters (Fig 1.a) \cite{RN9}. As a direct approach to reduce Gd dose, we also derive the 20\% low-dose CBV maps as illustrated in Figure \ref{Figure:Figure1}.b. For scans of tumor subjects, tumor masks are generated in addition to the brain masks using the Fuzzy-C-Means segmentation \cite{RN30}.

\subsection{Deep Learning Model}
\subsubsection{Model Architecture}
\begin{figure}
  \centering
  \includegraphics[width=0.8\textwidth]{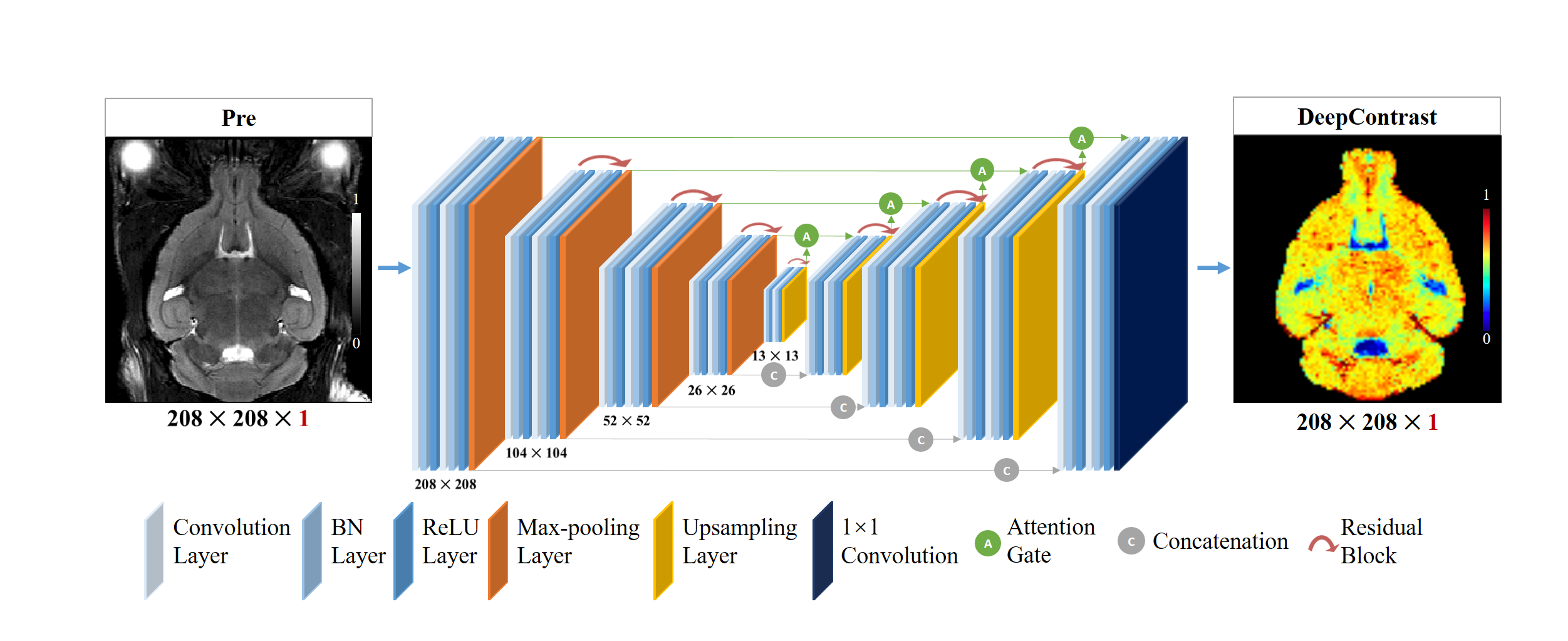} 
  \caption{Architecture visualization of the implemented ResAttU-Net. The proposed network is made of 5 encoding layers and 5 decoding layers. All the layers have residual connection induced; the decoding layers are implemented with an attention mechanism. The network takes only pre-scans as the single-channel input. Estimated 2D CBV maps with brain extraction are the output.}
  \label{Figure:Figure2}
\end{figure}

The deep learning architecture utilized in our study is the out-stand five-layer residual attention U-Net (ResAttU-Net) as illustrated in Figure \ref{Figure:Figure2}. This consists of a contraction path that encodes high-resolution data into low-resolution representations and an expansion path that decodes such encoded representations back to high-resolution images. We implemented our deep learning model using PyTorch framework with CUDA 10.0, 2 NVIDIA RTX 2080-TI GPUs and CentOS 6.

Both the encoder and decoder parts are based on the U-Net structure \cite{RN44} where each stage consists of two series of 3$\times$3 2D convolutions, batch normalization and rectified linear units (ReLU). In the encoder part, each stage is followed by 2$\times$2 max-pooling for down sampling, while for the decoder part, four 2$\times$2 upsampling layers convert low-resolution representation back to high resolution. Additionally, every stage of the decoder has concatenation from the encoder at the same level to give the model more accurate local information for assembling a precise prediction.

To further refine our network, as illustrated in Figure \ref{Figure:Figure2}, the residual blocks and the attention gates are incorporated into the U-Net architecture. The residual block can optimize the signal and gradient propagation within a network while preventing overfitting from being a problem \cite{RN45}. While the attention gate can significantly suppress feature responses in irrelevant background regions without needing to crop an ROI \cite{article14}.

\subsubsection{Training the Model}
We use ADAM as the optimizer with a learning rate of 10-3 and maximum 200 epochs with early stop. We used a batch size of 3 and randomize non-overlapping input images for training. Scans of 6 standalone mice were used for validation.

\subsubsection{Applying the Model}
We first apply the ResAttU-Net to derive the steady-state CBV maps in WT mice directly from their noncontrast pre-scans. In addition to demonstrating the ability to produce CBV in the normal brain tissue, we aimed to further verify its utility in the enhancement of pathology visibility and delineation of brain lesions as our second objective. For the first aim, 49 WT mice are used in the study, with a randomized 37-6-6 train-validation-test split. For the second aim, we randomly added 6 GBM mice scans to the training set and retrained the model. The performance of DeepContrast on tumor enhancement was tested on 4 GBM mice scans. 

\subsection{Evaluation and Statistical Analysis}
To evaluate the performance of DeepContrast, we used a  peak signal-to-noise ratio (PSNR) to assess the estimation error at the voxel level. Given the limitation of PSNR on capturing the perceptually relevant differences, we further calculated the structural similarity index (SSIM) to evaluate the accuracy of estimating a processed image on the structural level \cite{PSNR}. In addition, we conducted Pearson and Spearman correlation analysis to assess the linear and monotonic associations between the CBV ground truth and DeepContrast respectively. For the GBM study, besides the voxel level comparison, we utilized the Dice similarity coefficient and Hausdorff distance to compare the performance of DeepContrast and the CBV ground truth in tumor segmentation. 

\section{RESULTS}
\subsection{Performance of DeepContrast in Normal Brain CBV Mapping}
\begin{figure}
  \centering
  \includegraphics[width=0.8\textwidth]{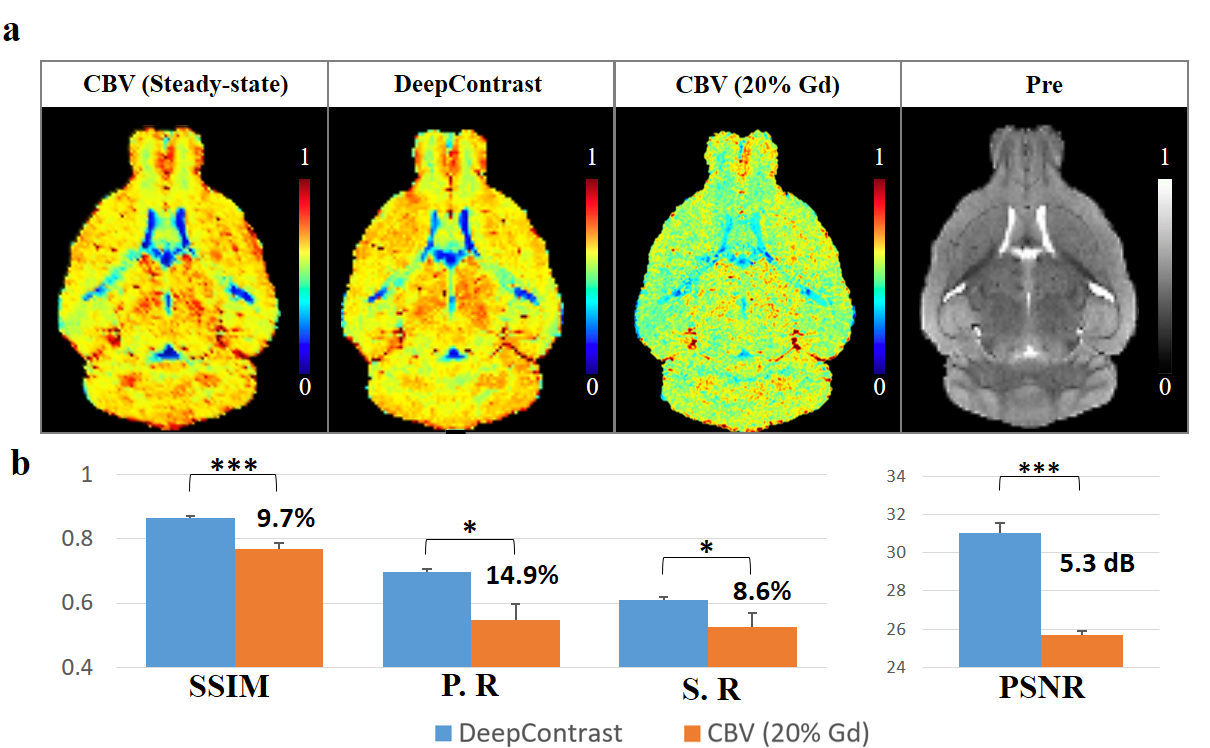} 
  \caption{Quantitative evaluation of the DeepContrast in the normal brain tissue. a) Steady-state CBV ground truth, DeepContrast, 20\% Gd CBV map and the Pre image of a representative mouse brain. b) Quantitative comparisons between the DeepContrast results and the 20\% Gd CBV.  DeepContrast prediction with input of only Pre image showed significant improvement compared to a 20\% low dose result. Statistical analyses used paired t-test; P.R and S.R are Pearson and Spearman Correlations respectively; Values denote mean $\pm$ S.E.M. $^{*}$P < 0.05, $^{**}$P < 0.01, $^{***}$P < 0.001. }
  \label{Figure:Figure3}
\end{figure}

Example of the DeepContrast prediction and the quantitative metrics of the standalone 6 testing subjects are shown in Figure \ref{Figure:Figure3}. DeepContrast captured the high contrast and fine details of small vessels with high similarity to the steady-state CBV ground truth in the normal brain tissue. Figure \ref{Figure:Figure3}.a shows the CBV ground truth, predicted result, 20\% low Gd dose CBV and noncontrast Pre image of the same mouse. The Pre image alone can provide promising prediction results that show strong enhancement compared to the 20\% low-dose CBV that are consistent with the steady-state CBV maps. Figure \ref{Figure:Figure3}.b shows the quantitative comparison of the DeepContrast and 20\% Gd CBV map. DeepContrast clearly improved the CBV contrast derived from 20\% Gd enhancement with a gain of 9.7\% SSIM (p < 0.001), an increase of 5.3 dB in PSNR (p < 0.001), an improvement of 14.9\% increase of P.R (Pearson Correlation, p < 0.05) and an 8.6\% increase in S.R (Spearman Correlation, p < 0.05).

\subsection{Performance of DeepContrast in GBM CBV Enhancement}
To experiment and evaluate the utility of our DeepContrast method in the enhancement of pathology visibility and delineation of brain lesions, we retrained the same ResAttU-Net architecture with additional MRI scans of the GBM mouse model. 

\begin{figure}
  \centering
  \includegraphics[width=0.8\textwidth]{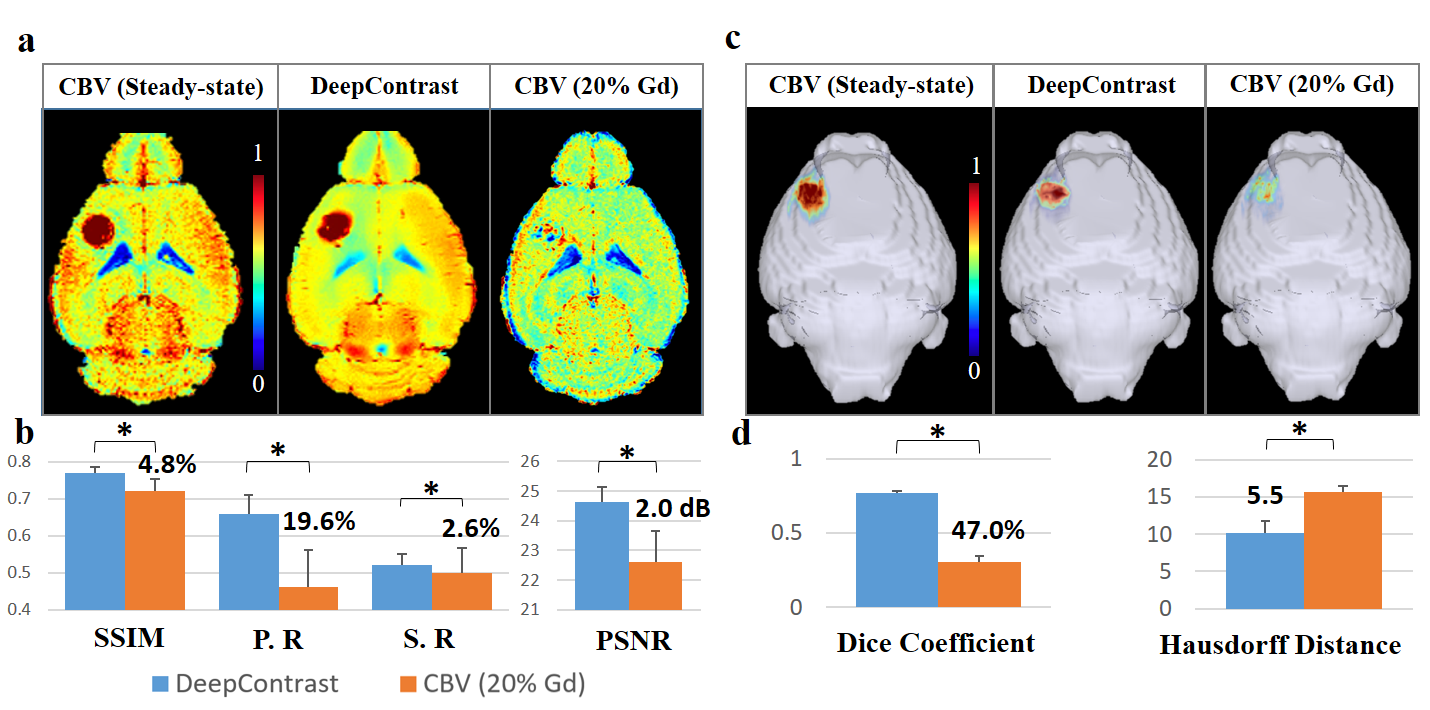} 
  \caption{DeepContrast performance for tumor enhancement and segmentation. a) The steady-state CBV, DeepContrast and 20\% low-dose CBV map of the same brain slice. The DeepContrast prediction with input of Pre only image shows significant contrast enhancement and tumor region prediction compared to 20\% low-dose CBV. b) Quantitative comparisons between the DeepContrast results and the 20\% low-dose CBV. c) 3D tumor region rendering figures of steady-state CBV, DeepContrast and 20 \% low-dose CBV map. The DeepContrast prediction with input of Pre only image shows similar enhanced contrast of tumor region that match the DeepContrast result. d) Quantitative comparisons (Dice coefficient and Hausdorff distance) between the DeepContrast results and the 20\% low-dose CBV. Statistical analyses were done using paired t-test; P.R and S.R are Pearson and Spearman Correlations respectively; Values denote mean $\pm$ S.E.M. $^{*}$P < 0.05, $^{**}$P < 0.01, $^{***}$P < 0.001.}
  \label{Figure:Figure4}
\end{figure}

Figure \ref{Figure:Figure4}.a-b show that the DeepContrast predicted results perform significantly better than the 20\% Gd CBV in both visual assessment and the quantitative evaluation. Figure \ref{Figure:Figure4}.a shows the predicted result of one slice of the same tumor subject. Compared with 20\% Gd CBV, DeepContrast generated from the T2W Pre scan shows similar contrast level to the ground truth. The consistency can be observed in the fine structure and significant contrast enhancement in both normal tissue and the tumor region. Figure \ref{Figure:Figure4}.b is the detailed quantitative metrics evaluation of 4 randomly selected testing tumor subjects. Compared with the 20\% Gd CBV, DeepContrast has a 4.8\% increase in SSIM (P < 0.05), a 2.0 dB increase in PSNR (P < 0.05), and a 19.6\% increase in P.R (Pearson Correlation, P < 0.05) and a 2.6\% increase in S.R (Spearman Correlation, P < 0.05). Figure \ref{Figure:Figure4} confirms that the DeepContrast predicted results perform significantly better than the 20\% Gd CBV in both visual assessment and the quantitative evaluation.

Figure \ref{Figure:Figure4}.c-d further confirms that the DeepContrast models with the input of Pre only image significantly outperformed the 20\% Gd for the tumor region. Figure \ref{Figure:Figure4}.c shows the 3D rendering results of CBV ground truth vs DeepContrast in the FCM segmented tumor region of a single tumor subject. Compared to low dose CBV, both the two DeepContrast models derived the tumor region more precisely with a similar contrast level to the ground truth. 

Figure \ref{Figure:Figure4}.d shows the quantitative comparison between the DeepContrast predictions and the 20\% Gd CBV of the tumor region. The DeepContrast performed significantly better than the 20\% Gd of the tumor region, with 47.0\% Dice coefficient increases (P < 0.05) and 5.5 pixels Hausdorff distance reductions (P < 0.05).

\section{DISCUSSION AND CONCLUSIONS}
Findings of Gd retention necessitate efforts to develop novel approaches to MRI that can decrease or even eliminate Gd exposure. In the past, there have been several attempts to develop noncontrast MRI sequences such as ASL, TOF, and VASO \cite{RN2}. While such sequences have been successful and applied clinically, Gd based imaging still offers an unparalleled level of information in comparison. For example, Gd is widely used for steady-state CBV fMRI imaging to map basal brain metabolism in both mice and humans \cite{RN12,RN16}. CBV has been proven to be tightly coupled to regional metabolism in healthy and diseased brains \cite{RN36,RN37} and is useful in studying cognitive aging \cite{RN36}, Alzheimer disease (AD) \cite{RN37} and tumors \cite{RN38}. Gd-enhanced CBV is also well suited to detect GBM-related regional hyperactive angiogenesis and blood-brain-barrier (BBB) leakage which has become a potential imaging biomarker for GBM detection and grading \cite{RN23,RN24,RN25,RN26}.  

The question remains whether it is feasible to efficiently gather Gd-based information directly from noncontrast MRI scans to ensure patient safety and limit Gd exposure. Removing the Gd contrast entirely while retaining diagnostic information could have a large impact on both patient well-being and reduction of imaging time and costs. Recently, 3D Bayesian U-Net was applied to a dataset from patients with brain tumors and healthy subjects to predict contrast enhancement from a comprehensive multiparametric MRI protocol including T1w, T2w, T2w fluid-attenuated inversion recovery, diffusion-weighted imaging, and susceptibility-weighted imaging, all acquired without any Gd injections \cite{Can}. This method was limited in its inability to predict presence of small vessels and an inevitably long scan time likely to exhaust patients due to utilization of 10 multiparametric MRI scans as its input. With these limitations in mind, DeepContrast was developed to rely solely on information extracted from the most commonly acquired structural MRI scans. Compared to the 3D Bayesian U-Net, our approach provides several improvements to predicting contrast enhancement. First, we demonstrate that Gd contrast in brain MRI can be directly derived from a single noncontrast T2W MRI in both normal brain tissue and brain lesion. High PSN, high SSIM and high correlations between the DeepContrast and the Gd-enhanced ground truth suggest that our proposed method does not cause significant quality degradation. High accuracy of GBM segmentations suggest DeepContrast is capable of detecting GBM with performance similar to full-dose Gd-enhanced scans. Furthermore, our deep learning method is based on a hybrid deep residual attention-aware network and is the first network to use an attention residual mechanism to process brain MRI scans. The basic architecture of the proposed network is a 2D U-Net that extracts contextual information combining low-level feature maps with high-level ones. Attention modules are stacked such that the attention-aware features change adaptively in the deeper layers of the network. This is made possible by residual learning. Finally, our method is capable of producing CBV mapping of small vessels with high fidelity. 

While our method of DeepContrast enhancement shows promising results, there are several limitations that must be considered. Regarding application of the network to GBM mouse models, our training dataset lacking longitudinal scans to capture GBM contrast at different tumor stages. To improve the robustness and accuracy of DeepContrast for GBM enhancement, we could enrich our current training dataset by adding MRI scans of mice injected with GBM cells at various stages and different locations. Beyond diversifying the dataset, to deal with the intensity difference within and across subjects, recent advancements in intensity standardization / normalization would arguably benefit the estimation of the DeepContrast enhancement when  translating our proposed model to human data \cite{RN49}. An additional factor and potential limitation to consider is loss function parameters. MSE was chosen as the cost function in our current model, but the training strategy could be further improved by adding other evaluation parameters to the loss function. Overall, future works will incorporate larger datasets, enhanced image pre-processing, and model advancement to optimize both the network and training pipeline. 

Despite aforementioned limitations, our proposed method currently can be used to generate Gd contrast in brain MRI directly from T2W MRI scans with complete omission of GBCAs. DeepContrast is a promising technique with the potential to offer benefits to patient care and the healthcare system through reduction of Gd exposure, scan time, and cost. Of note, our work should be regarded as a proof-of-concept study in animal models; future human studies will be required to validate its clinical utility.
\section*{Acknowledgement}
This study was funded by the Seed Grant Program and Technical Development Grant Program and Matheson Foundation (UR010590). This study was performed at the Zuckerman Mind Brain Behavior Institute MRI Platform, a shared resource.

\bibliographystyle{unsrt}  
\bibliography{DeepCBV.bib} 

\begin{thebibliography}{10}

\bibitem{RN10}
Scott~A. Small, Scott~A. Schobel, Richard~B. Buxton, Menno~P. Witter, and
  Carol~A. Barnes.
\newblock A pathophysiological framework of hippocampal dysfunction in ageing
  and disease.
\newblock {\em Nature reviews. Neuroscience}, 12(10):585--601, 2011.

\bibitem{RN12}
Usman~A. Khan, Li~Liu, Frank~A. Provenzano, Diego~E. Berman, Caterina~P.
  Profaci, Richard Sloan, Richard Mayeux, Karen~E. Duff, and Scott~A. Small.
\newblock Molecular drivers and cortical spread of lateral entorhinal cortex
  dysfunction in preclinical alzheimer's disease.
\newblock {\em Nature neuroscience}, 17(2):304--311, 2014.

\bibitem{RN11}
Scott~A. Schobel, Nashid~H. Chaudhury, Usman~A. Khan, Beatriz Paniagua,
  Martin~A. Styner, Iris Asllani, Benjamin~P. Inbar, Cheryl~M. Corcoran,
  Jeffrey~A. Lieberman, Holly Moore, and Scott~A. Small.
\newblock Imaging patients with psychosis and a mouse model establishes a
  spreading pattern of hippocampal dysfunction and implicates glutamate as a
  driver.
\newblock {\em Neuron}, 78(1):81--93, 2013.

\bibitem{RN16}
Adam~M. Brickman, Usman~A. Khan, Frank~A. Provenzano, Lok-Kin Yeung, Wendy
  Suzuki, Hagen Schroeter, Melanie Wall, Richard~P. Sloan, and Scott~A. Small.
\newblock Enhancing dentate gyrus function with dietary flavanols improves
  cognition in older adults.
\newblock {\em Nature neuroscience}, 17(12):1798--1803, 2014.

\bibitem{RN5}
Tyler~E. Smith, Andrew Steven, and Bridget~A. Bagert.
\newblock Gadolinium deposition in neurology clinical practice.
\newblock {\em The Ochsner journal}, 19(1):17--25, 2019.

\bibitem{RN40}
{FDA}.
\newblock {FDA} drug safety communication: {FDA} warns that gadolinium-based
  contrast agents (gbcas) are retained in the body; requires new class
  warnings, 2018.

\bibitem{Gong}
Enhao Gong, John Pauly, Max Wintermark, and Greg Zaharchuk.
\newblock Deep learning enables reduced gadolinium dose for contrast-enhanced
  brain mri: Deep learning reduces gadolinium dose.
\newblock {\em Journal of Magnetic Resonance Imaging}, 48, 02 2018.

\bibitem{Can}
Jens Kleesiek, Jan Morshuis, Fabian Isensee, Katerina Deike-Hofmann, Daniel
  Paech, Philipp Kickingereder, Ullrich Köthe, Carsten Rother, Michael
  Forsting, Wolfgang Wick, Martin Bendszus, Heinz-Peter Schlemmer, and
  Alexander Radbruch.
\newblock Can virtual contrast enhancement in brain mri replace gadolinium?: A
  feasibility study.
\newblock {\em Investigative Radiology}, 54:1, 07 2019.

\bibitem{Lei}
Liang Lei, Adam Sonabend, Paolo Guarnieri, Craig Soderquist, Thomas Ludwig,
  Steven Rosenfeld, Jeffrey Bruce, and Peter Canoll.
\newblock Glioblastoma models reveal the connection between adult glial
  progenitors and the proneural phenotype.
\newblock {\em PloS one}, 6:e20041, 05 2011.

\bibitem{RN9}
Herman Moreno, Fan Hua, Truman Brown, and Scott Small.
\newblock Longitudinal mapping of mouse cerebral blood volume with mri.
\newblock {\em NMR in Biomedicine}, 19(5):535--543, 2006.

\bibitem{RN30}
Yogita Dubey and Milind Mushrif.
\newblock Fcm clustering algorithms for segmentation of brain mr images.
\newblock {\em Advances in Fuzzy Systems}, 2016:1--14, 2016.

\bibitem{RN44}
Olaf Ronneberger, Philipp Fischer, and Thomas Brox.
\newblock U-net: Convolutional networks for biomedical image segmentation.
\newblock {\em CoRR}, abs/1505.04597, 2015.

\bibitem{RN45}
Kaiming He, Xiangyu Zhang, Shaoqing Ren, and Jian Sun.
\newblock Deep residual learning for image recognition.
\newblock {\em CoRR}, abs/1512.03385, 2015.

\bibitem{article14}
Ozan Oktay, Jo~Schlemper, Lo{\"{\i}}c~Le Folgoc, Matthew C.~H. Lee, Mattias~P.
  Heinrich, Kazunari Misawa, Kensaku Mori, Steven~G. McDonagh, Nils~Y.
  Hammerla, Bernhard Kainz, Ben Glocker, and Daniel Rueckert.
\newblock Attention u-net: Learning where to look for the pancreas.
\newblock {\em CoRR}, abs/1804.03999, 2018.

\bibitem{PSNR}
A.~{Horé} and D.~{Ziou}.
\newblock Image quality metrics: Psnr vs. ssim.
\newblock In {\em 2010 20th International Conference on Pattern Recognition},
  pages 2366--2369, Aug 2010.

\bibitem{RN2}
Jessica Lohrke, Thomas Frenzel, Jan Endrikat, Filipe~Caseiro Alves, Thomas~M.
  Grist, Meng Law, Jeong~Min Lee, Tim Leiner, Kun-Cheng Li, Konstantin
  Nikolaou, Martin~R. Prince, Hans~H. Schild, Jeffrey~C. Weinreb, Kohki
  Yoshikawa, and Hubertus Pietsch.
\newblock 25 years of contrast-enhanced mri: Developments, current challenges
  and future perspectives.
\newblock {\em Advances in Therapy}, 33(1):1--28, 2016.

\bibitem{RN36}
William Jagust.
\newblock Vulnerable neural systems and the borderland of brain aging and
  neurodegeneration.
\newblock {\em Neuron}, 77(2):219--234, 2013.

\bibitem{RN37}
Jinsoo Uh, Kelly Lewis-Amezcua, Kristin Martin-Cook, Yamei Cheng, Myron Weiner,
  Ramon Diaz-Arrastia, Sr. Devous, Michael, Dinggang Shen, and Hanzhang Lu.
\newblock Cerebral blood volume in alzheimer's disease and correlation with
  tissue structural integrity.
\newblock {\em Neurobiology of aging}, 31(12):2038--2046, 2010.

\bibitem{RN38}
Diego J~Covarrubias, Bruce Rosen, and Michael Lev.
\newblock Dynamic magnetic resonance perfusion imaging of brain tumors.
\newblock {\em The oncologist}, 9:528--37, 2004.

\bibitem{RN23}
Małgorzata Neska-Matuszewska, Joanna Bladowska, Marek Sąsiadek, and Anna
  Zimny.
\newblock Differentiation of glioblastoma multiforme, metastases and primary
  central nervous system lymphomas using multiparametric perfusion and
  diffusion mr imaging of a tumor core and a peritumoral zone-searching for a
  practical approach.
\newblock {\em PloS one}, 13(1):e0191341--e0191341, 2018.

\bibitem{RN24}
Junfeng Zhang, Heng Liu, Haipeng Tong, Sumei Wang, Yizeng Yang, Gang Liu, and
  Weiguo Zhang.
\newblock Clinical applications of contrast-enhanced perfusion mri techniques
  in gliomas: Recent advances and current challenges.
\newblock {\em Contrast media \& molecular imaging}, 2017:7064120--7064120,
  2017.

\bibitem{RN25}
Meng Law, Stanley Yang, James~S. Babb, Edmond~A. Knopp, John~G. Golfinos, David
  Zagzag, and Glyn Johnson.
\newblock Comparison of cerebral blood volume and vascular permeability from
  dynamic susceptibility contrast-enhanced perfusion mr imaging with glioma
  grade.
\newblock {\em American Journal of Neuroradiology}, 25(5):746, 2004.

\bibitem{RN26}
Meng Law, Stanley Yang, Hao Wang, James~S. Babb, Glyn Johnson, Soonmee Cha,
  Edmond~A. Knopp, and David Zagzag.
\newblock Glioma grading: Sensitivity, specificity, and predictive values of
  perfusion mr imaging and proton mr spectroscopic imaging compared with
  conventional mr imaging.
\newblock {\em American Journal of Neuroradiology}, 24(10):1989, 2003.

\bibitem{RN49}
Attila Simk{\'o}, Tommy L{\"o}fstedt, Anders Garpebring, Tufve Nyholm, and
  Joakim Jonsson.
\newblock A generalized network for {\{}mri{\}} intensity normalization.
\newblock In {\em International Conference on Medical Imaging with Deep
  Learning -- Extended Abstract Track}, London, United Kingdom, 08--10 Jul
  2019.

\end{thebibliography}
\end{document}